\documentclass[prc,showpacs,nofootinbib]{revtex4}

\newcommand{\be}{\begin{equation}}
\newcommand{\ee}{\end{equation}}
\usepackage{graphicx}
\usepackage{psfig}
\usepackage{epsfig}
\usepackage{amsmath}
\begin{document}
\title{Covariant propagator of the Rarita-Schwinger field in the nuclear medium}
\author{C.L. Korpa}
\affiliation{Department of Theoretical Physics, University of P\'ecs,
H-7624 Pecs, Hungary}
\author{A.E.L. Dieperink}
\affiliation {KVI, Zernikelaan 25, NL-9747AA Groningen, The
Netherlands}
\date{\today}
\begin{abstract}
A formalism for representing the fully relativistic propagator of the
Rarita-Schwinger field 
in the nuclear medium is developed.
Using a convenient basis for expanding the propagator 
it is shown that it can be represented by 40 energy
and momentum dependent quantities which can be decomposed into an 2$\times$2 and
an 6$\times$6 matrix. In this way calculations reduce to matrix multiplication. 
Using the presented formalism the full relativistically covariant contribution 
of the pion-nucleon loop to the 
isobar self energy and propagator in isospin-symmetric spin-saturated 
nuclear medium is computed. Utilizing this propagator the 
photoabsorption cross-section on in-medium nucleons in the isobar region is
calculated and
the result compared with experimental data.
\end{abstract}
\pacs{25.20.Dc,24.10.Jv,21.65.+f}
 \keywords{Isobar, relativistic, self energy, photoabsorption}
\maketitle
\section{Introduction}
The spin-3/2 isospin-3/2 $\Delta$(1232) baryon (in the following refered to as the
isobar and denoted by $\Delta$) couples strongly to the nucleon
and pion and plays an important role in nuclear processes with 
excitation energy of a few hundred MeV. 
This raisies the issue of the propagation of a spin-3/2 particle in the nuclear
medium. 
The spin-3/2 particle is usually described by using the Rarita-Schwinger
field \cite{rarita41}, which consists of four, Lorentz-vector indexed, Dirac spinors. 
A free theory can be formulated in which the number of degrees of freedom is
as a consequence of the equations of motion reduced to 
the required 8 (for a complex field). However, for realistic description of spin-3/2 resonances
one needs to introduce interactions taking into account their decay channels.

Introducing interactions of general form into the above free-field theory 
introduces also spin-1/2 components into the propagator. This type of
approach was used quite extensively in the past for the delta baryon and 
was phenomenologically successful (see, for example, Ref.~\cite{davidson91}).
Recently it has been argued \cite{pascalutsa98,pascalutsa99} that an interaction vertex 
which preserves the correct number of degrees of freedom for the spin-3/2 particle 
should be used.  
It has been also  shown \cite{pascalutsa01} that the two types of
couplings can be transformed into each other by a redefinition of the field
describing the spin-3/2 particle, at the expense of new contact interactions appearing (not
containing the spin-3/2 field) in the lagrangian. These contact interactions may be associated
with other spin-1/2 fields in the theory.

The propagator we are interested in is defined by
\be
G^{\mu\nu}(p)=i\int d^4 x e^{i p\cdot x}\,\langle 0|\text{T}\,\Psi^\mu(x) \bar\Psi^\nu(0)|0
\rangle ,
\label{deltaprop}
\ee
where $\Psi^\mu$ is the Rarita-Schwinger field and $|0\rangle$ can denote either the 
vacuum or the nuclear medium. In vacuum this
propagator (or the self energy, which is given by the amputated diagrams of
(\ref{deltaprop}))
can be expanded in terms of 10 Lorentz-scalar
functions of $p^2$. This number can easily be understood if we count
the number of terms corresponding to the Lorentz structure given by Eq.~(\ref{deltaprop}).
Indeed, the two vector indices can be represented by the 5 terms $g^{\mu\nu}, p^\mu p^\nu,
p^\mu \gamma^\nu, \gamma^\mu p^\nu, \gamma^\mu \gamma^\nu$ ($\gamma^\mu$ are the
Dirac matrices); these are multiplied by the 4$\times$4 
unit matrix, $\openone$, or the contraction $p\hspace{-1.7mm}/ \equiv 
p_\alpha \gamma^\alpha$.

In the nuclear medium the propagator acquires more independent components than in
vacuum, since the presence of the (rotationally symmetric) medium means that 
preserving the Lorentz-covariant description requires 
introducing another four-vector, namely the medium's four velocity. This quadruples 
the number of terms the propagator can be expanded in (see 
section \ref{rel-delta} for details). 
To our knowledge a complete calculation involving all these terms has not
been carried out and 
either a nonrelativistic
approach (for example as in Ref.~\cite{korpa95}), a description based on
combining off-shell and on-shell behavior \cite{dejong92},
or a treatment based on the assumption that the isobar 
propagator is proportional to the spin-3/2 projector 
\cite{xia94,vandaele02} was used. The implications of this latter approach
we discuss in some detail in the subsection \ref{isobar}.
Our aim is to provide a general scheme for a fully
Lorentz-covariant treatment of the spin-3/2 particle's in-medium propagation, analitically
separating all its components. This means introducing a convenient basis
for the expansion of the self energy and the propagator, which makes subsequent
calculations practical for arbitrary vertices involving the Rarita-Schwinger field.

In the first part of section \ref{rel-delta} we develop the general formalism
for a covariant treatment of the spin-3/2 field in the nuclear medium, while in the
second part we provide numerical results for the in-medium delta propagator.
In section \ref{photo} we calculate the total cross section for the nuclear
photoabsorption and compare the results with observations.
\section{Relativistic propagator of the in-medium Rarita-Schwinger field}
\label{rel-delta}

\subsection{Formalism}
To exhibit the general Lorentz-structure of the spin-3/2 particle's propagator and 
self energy
in the medium, we observe
that in the presence of rotationally symmetric nuclear medium the latter's state
of motion is given by its four-velocity, $u^\mu\equiv 1/\sqrt{1-\vec v^2/c^2}
(1,\vec v/c)$. This implies that in addition to the 5 objects with two
Lorentz-vector indices in vacuum (see the preceding section),
we get 5 more: $u^\mu p^\nu, p^\mu u^\nu, u^\mu\gamma
^\nu, \gamma^\mu u^\nu, u^\mu u^\nu$. Also, the 4$\times$4 matrices 
$u\hspace{-1.9mm}/$ and $p\hspace{-1.7mm}/ u\hspace{-1.9mm}/$, in addition
to $\openone$ and $p\hspace{-1.7mm}/$ appear,
producing in all 40 terms of the basis. 

While in vacuum one may argue that the definite spin of the particle 
restricts the form of the propagator and decreases the number of the nonzero
expansion coefficients, in medium such an argument is meaningless. The reason is
that, even in a rotationally symmetric medium, the spin of a particle moving
with respect to the medium is not a good quantum number, since the momentum
of the particle (in the rest frame of the medium) defines a preferred direction
and only rotations about that axis are a symmetry transformation. 
This means that only the projection
of the angular momentum on the momentum, i.e.\ helicity, is a good quantum 
number. As a consequence, not only spin-3/2 and spin-1/2 components, but also
those corresponding to mixing of these spins should in general appear in the 
propagator, with the restriction that helicity-1/2 and helicity-3/2
states do not mix. We remark that the sign of the helicity does not matter, since
parity is a symmetry operation.

A convenient basis containing the 40 terms mentioned above has actually been
constructed in Ref.~\cite{lutz02}, when an in-medium generalization of the partial-wave
expansion has been introduced. In that case a basis for s-, p- and d-waves
with $J=1/2$ and $J=3/2$ (including their mixing) was obtained, containing 68 terms. 
Of these only 40 terms 
are relevant for the present case, since the rest does not have two Lorentz-vector
indices.

To introduce the relevant basis we recapitulate the definition and basic properties
of its building blocks. 
First, the terms $P_\pm$ and $U_\pm$ are introduced\footnote{In Ref.\ \cite{lutz02}
due to typographical errors in Eq.\ (A1) a minus sign is missing in the defining
expression for $U_\pm$, and also the sign of the terms containing $U_+(w,u)+U_-(w,u)$
in equations defining $R_\mu$ and $L_\mu$ should be changed.} 
providing the 4$\times$4 matrix
structure, by the relations:
\be
P_\pm(p) = \frac{1}{2}\left( 1\pm \frac{p\hspace{-1.7mm}/}{\sqrt{p^2}}\right)\, ,\quad 
U_\pm (p,u)=-P_\pm(p) \,\frac{i\,\gamma \cdot u}{\sqrt{(p\cdot u)^2/p^2-1}}\,P_\mp(p)\;.
\ee
They have the following multiplication properties:
\begin{eqnarray}
&&P_\pm P_\pm=P_\pm=U_\pm U_\mp\,,\quad  P_\pm P_\mp=0=U_\pm U_\pm\,,\nonumber\\
&&P_\pm U_\pm=U_\pm=U_\pm P_\mp\,,\quad  P_\pm U_\mp=0=U_\pm P_\pm\,.
\end{eqnarray}
One can observe that on-shell the $P_\pm(p)$ are the positive- and negative-energy
projectors.

To exhibit the remaining rotational symmetry in the medium, leading to the conservation of
helicity, it is advantageous to use, apart from $p^\mu$, the following four 
quantities \cite{lutz02}: 
\begin{eqnarray}
&&V_\mu (p)=\frac{1}{\sqrt{3}}\,\Big( \gamma_\mu -\frac{p\hspace{-1.7mm}/}{p^2}\,p_\mu \Big)
\;,\quad X_\mu(p,u)= \frac{(p\cdot u)\,p_\mu-p^2\,u_\mu}{p^2\,\sqrt{(p \cdot u)^2/p^2-1}} \;,
\nonumber\\
&&R_\mu (p,u) = \frac{1}{\sqrt{2}}\,\Big( U_+(p,u)+U_-(p,u)\Big)\,V_\mu(p)-i\,
\sqrt{\frac{3}{2}}\,X_\mu(p,u)
\, , \quad 
\nonumber\\
&& L_\mu(p,u) =\frac{1}{\sqrt{2}}\,V_\mu(p)\, \Big( U_+(p,u)+U_-(p,u)\Big) -i\,
\sqrt{\frac{3}{2}}\,X_\mu(p,u) \;,
\label{def-basic}
\end{eqnarray}
While they are not all independent, they are constructed in such a 
way to make the basis orthogonal, see Eq.\ (\ref{orthog}) below, and are motivated by
the helicity basis for the partial-wave expansion \cite{lutz02}. 

It turns out that the basis can be constructed in such a way that
the multiplication algebra of its
40 objects separates into two subalgebras, one containing 4 and the other 
36 terms. 
The existence of two separate subalgebras of the expansion basis is 
explained by the helicity conservation in the medium. The 4 terms of the 
first subalgebra can be cast in the form of a 2$\times$2 matrix, which
corresponds to helicity-3/2 components, containing the positive-energy
and negative-energy states (and their mixing). The remaining 36 terms, 
suitably grouped in a
6$\times$6 matrix, describe the helicity-1/2 components. They come from
the spin-3/2 part of the Rarita-Schwinger field and the two spin-1/2 
sectors present in that field, and positive and negative energy 
components for each of the above mentioned spin sectors.

The
4 terms of the first subalgebra are given by $Q^{\mu\nu}_{[ij]}$ with $i,j=1,2$, defined
as follows:
\begin{eqnarray}
&& Q_{[11]}^{\mu \nu } =\Big( g^{\mu \nu}-\hat p^\mu\,\hat p^\nu \Big)
\,P_+ - V^\mu\,P_-\,V^\nu -L^\mu\,P_+\,R^\nu \;, 
\nonumber\\
&& Q_{[22]}^{\mu \nu } =\Big( g^{\mu \nu}-\hat p^\mu\,\hat p^\nu \Big)
\,P_- - V^\mu\,P_+\,V^\nu -L^\mu\,P_-\,R^\nu \;, 
\nonumber\\
&& Q_{[12]}^{\mu \nu }  = \Big(g^{\mu \nu}-\hat p^\mu\,\hat p^\nu \Big)\,U_+ 
+{\textstyle{\frac{1}{3}}}\,V^\mu\,U_-\,V^\nu 
\nonumber\\
&&\qquad +{\textstyle{\frac{\sqrt{8}}{3}}}\,
\Big( L^\mu\,P_+\,V^\nu +V^\mu\,P_-\,R^\nu \Big) -{\textstyle{\frac{1}{3}}}\,L^\mu\,U_+\,R^\nu\;,
\nonumber\\
&&Q_{[21]}^{\mu \nu } = \Big(g^{\mu \nu}-\hat p^\mu\,\hat p^\nu\Big)\,U_- 
+{\textstyle{\frac{1}{3}}}\,V^\mu\,U_+\,V^\nu 
\nonumber\\
&&\qquad +{\textstyle{\frac{\sqrt{8}}{3}}}\,
\Big( L^\mu\,P_-\,V^\nu +V^\mu\,P_+\,R^\nu \Big)-{\textstyle{\frac{1}{3}}}\,L^\mu\,U_-\,R^\nu\;,
\label{qs}
\end{eqnarray}
where $\hat p_\mu = p_\mu /\sqrt{p^2} $.

The 36 components $P^{\mu\nu}_{[ij]}$ of the second subalgebra are given by the relation:
\be
P^{\mu\nu}_{[ij]}=P^\mu _i \bar{P}^\nu_j,\qquad i,j=1,\cdots ,6\,,
\label{ps}
\ee
where:
\begin{eqnarray}
&&\begin{array}{ll}
P^\mu_{1} = V_\mu \,P_+ \,,  & 
\bar P^\mu_{1} = P_+\,V^\mu \;,  \\
P^\mu_{2} = V_\mu \,U_- \;,  &
\bar P^\mu_{2} = U_+\,V^\mu\;,  \\
P^\mu_{3} = \hat p_\mu \,P_+ \;,  &
\bar P^\mu_{3} = P_+\,\hat p^\mu \;, \\
P^\mu_{4} = \hat p_\mu \,U_- \;,  & 
\bar P^\mu_{4} = U_+\,\hat p^\mu\;, \\
P^\mu_{5} =   L_\mu \,P_+ \;,  & 
\bar P^\mu_{5} = P_+\,  R^\mu \;, \\
P^\mu_{6} =   L_\mu \,U_- \;,  &
\bar P^\mu_{6} = U_+\,  R^\mu\;.
\end{array}\label{ps1}
\end{eqnarray}
The $Q^{\mu\nu}_{[ij]}$ and $P^{\mu\nu}_{[ij]}$ satisfy the following relations:
\begin{eqnarray}
&& Q_{[ik]}^{\mu \alpha }\,g_{\alpha \beta}\,P_{[lj]}^{\beta \nu }
= 0 = P_{[ik]}^{\mu \alpha }\,g_{\alpha \beta}\,Q_{[lj]}^{\beta \nu }
\;, \nonumber\\
&& Q_{[ik]}^{\mu \alpha }\,g_{\alpha \beta}\,Q_{[lj]}^{\beta \nu }
= \delta_{kl}\,Q_{[ij]}^{\mu \nu} \;,\quad 
P_{[ik]}^{\mu \alpha }\,g_{\alpha \beta}\,P_{[lj]}^{\beta \nu }
= \delta_{kl}\,P_{[ij]}^{\mu \nu}\;.
\label{orthog}
\end{eqnarray}

The free and the dressed propagator, as well as the self energy 
of the Rarita-Schwinger field can be expanded in the basis
spanned by $Q^{\mu\nu}_{[ij]}$ and $P^{\mu\nu}_{[ij]}$:
\begin{subequations}
\label{exps}
\begin{eqnarray}
&&
G_0^{\mu\nu}(p)=\sum_{i,j=1}^2 Q^{\mu\nu}_{[ij]} \,g^{(Q)}_{[ij]}(p)+
\sum_{i,j=1}^6 P^{\mu\nu}_{[ij]}\, g^{(P)}_{[ij]}(p),\label{expa}\\
&& G^{\mu\nu}(p,u)=\sum_{i,j=1}^2 Q^{\mu\nu}_{[ij]}\, G^{(Q)}_{[ij]}(p,u)+
\sum_{i,j=1}^6 P^{\mu\nu}_{[ij]} \,G^{(P)}_{[ij]}(p,u),\label{expb}\\
&& \Sigma^{\mu\nu}(p,u)=\sum_{i,j=1}^2 Q^{\mu\nu}_{[ij]}\, \sigma^{(Q)}_{[ij]}(p,u)+
\sum_{i,j=1}^6 P^{\mu\nu}_{[ij]} \,\sigma^{(P)}_{[ij]}(p,u),\label{expc}
\end{eqnarray}
\end{subequations}
where the dependence of $Q^{\mu\nu}_{[ij]}$ and $P^{\mu\nu}_{[ij]}$ on 
$p$ and $u$ is not shown. Using the orthogonality relations (\ref{orthog})
the Dyson equation
\be
G^{\mu\nu}=G_0^{\mu\nu}+G_0^{\mu\alpha}\,g_{\alpha\beta}\,\Sigma^{\beta\gamma}\,
g_{\gamma\delta}\,G^{\delta\nu}
\ee
simply becomes a matrix equation, where the matrices are constructed from the
corresponding expansion coefficients $c_{[ij]}$, with $i$ being the row index,
$j$ the column index:
\begin{eqnarray}
&&
G^{(Q)}_{[ij]}=g^{(Q)}_{[ij]}+\sum_{k,\ell=1}^2 g^{(Q)}_{[ik]}\,\sigma^{(Q)}_{[k\ell]}\,
G^{(Q)}_{[\ell j]},\nonumber\\
&&
G^{(P)}_{[ij]}=g^{(P)}_{[ij]}+\sum_{k,\ell=1}^6 g^{(P)}_{[ik]}\,\sigma^{(P)}_{[k\ell]}\,
G^{(P)}_{[\ell j]}.
\end{eqnarray}
This means we can simply solve for the expansion functions of the propagator:
\be
\tilde G^{(A)}=(\tilde g^{(A)-1}-\tilde \sigma^{(A)})^{-1},
\ee
with $A=Q,P$ and $\tilde G$ denoting a matrix with elements $G_{[ij]}$.

\subsection{Relativistic propagator of the isobar}\label{isobar}
We start from the free isobar propagator in its standard form
\be
G_0^{\mu\nu}(p)=\frac{p\hspace{-1.6mm}/ +M_\Delta}{p^2-M^2_\Delta+
i\varepsilon}
\left[ g^{\mu\nu}-\frac{\gamma^\mu\gamma^\nu}{3}
-\frac{2p^\mu p^\nu}{3M^2_\Delta}
+\frac{p^\mu\gamma^\nu-p^\nu\gamma^\mu}{3M_\Delta}
\right].\label{freeprop}
\ee
The expansion coefficients of its inverse can be simply calculated by 
matrix inversion and are given in Appendix \ref{apA}.

The self energy of the isobar close to its on-shell energy is dominated
by the pion-nucleon loop and for the relevant
pion-nucleon-isobar coupling we use the expression
\be
{\mathcal{L}}_{\pi\text{N}\Delta}=
g_{\pi\text{N}\Delta} \partial_\alpha \pi \bar{\Delta}_\beta
(g^{\alpha\beta}+a \gamma^\beta \gamma^\alpha)N + \text{h.c.}
\label{pind}
\ee
An
explicit computation in vacuum based on the coupling (\ref{pind}) showed 
\cite{korpa97} that by introducing a pion-nucleon-isobar form factor 
depending on the square of isobar's 4-momentum (since the pion and the nucleon are
on-shell), an excellent fit to the pion-nucleon scattering
phase shift in the isospin-3/2 spin-3/2 channel is possible, up to the pion
laboratory momentum of 500 MeV, by adjusting the coupling, the isobar's bare mass and
the cut-off in the form factor. The off-shell
parameter $a$, as expected, does not play a significant role in the energy 
range of interest.
Also, that calculation
showed that the spin-1/2 components of the isobar spectral function are about two orders
of magnitude smaller 
than the spin-3/2 ones in the resonance region.
The Rarita-Schwinger propagator in vacuum
was recently discussed in Ref.~\cite{kaloshin04}, where some elements of
the basis used in the present approach were introduced (of course, without the terms 
relevant for the nuclear medium).

For the in-medium calculation of the isobar self energy we also use the interaction lagrangian
(\ref{pind}), but with the off-shell parameter $a$ put to zero. This is
mainly done for simplicity and is motivated by the observation of 
Ref.\ \cite{korpa97} that this parameter has a very small effect on the
isobar propagator (in vacuum) in the resonance region, i.e.\ not far off-shell.
The coefficients of the self-energy expansion are
given in Appendix \ref{apA}, Eq.\ (\ref{sigmaexp}), together with the 
definition of the pion-nucleon  
loop integrals appearing in those functions. These loop integrals correspond
to the ones defined in Ref.\ \cite{lutz02} for the case of kaon-nucleon scattering.

For a comparison with the result of the non-relativistic approach we 
calculated the isobar propagator using the pion spectral function and
the $\pi\textrm{N}\Delta$ coupling and form factor of Ref.\ \cite{korpa95}.
We observe that two entries in the propagator, one in $\tilde G^{(Q)}$ and 
one in $\tilde G^{(P)}$,
are much larger than any of the remainder, which brings us again to the
interpretation of these matrix elements. As discussed in the previous subsection, 
the appearence of the block
diagonal structure, i.e.\ matrices with superscripts $Q$ and $P$, is
a reflection of the fact that helicity is still a good quantum number and
consequently different helicity states do not mix. The 2$\times$2 matrix
contains the helicity-3/2 components (see also Ref.\ \cite{lutz02}) of the
positive and negative energy states and their mixing. The [11] element of this
matrix is much larger (for positive energies) than the others and is 
referred to as the helicity-3/2 component of the in-medium isobar. We repeat
that the projector $P_+$ appearing in this term when on-shell, i.e.\ for
$p^2=M^2$, is the positive energy projector. 

The matrix $\tilde G^{(P)}$ contains the helicity-1/2 components, of both
spin-1/2 and spin-3/2 (which mix), and positive-energy and negative-energy components.
The element [55] of this matrix is the dominant one and is identified with the
spin-3/2 term (also positive energy). The appearance of the 6$\times$6 matrix
structure, and not a 4$\times$4 one, is a consequence of the number of degrees
of freedom of the Rarita-Schwinger field, which means that there are two
spin-1/2 sectors in that field.

\begin{figure}
\centering{
\includegraphics[width=0.6\textwidth] {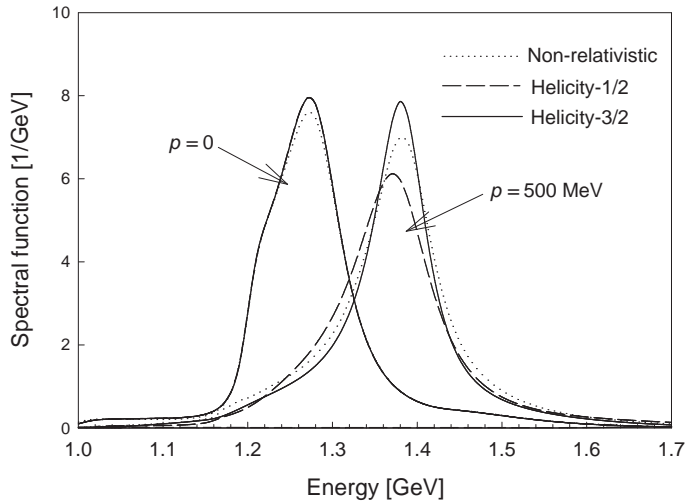}
\caption{\label{fig1}The isobar spectral function in the nuclear medium at
saturation density and momentum $|\vec p|=0$ and $|\vec p|=500\;$MeV. 
The pion spectral function and the $\pi\textrm{N}\Delta$
form factor, as well as the dotted line result for the isobar spectral function 
are from a non-relativistic calculation of Ref.\cite{korpa95}. At zero momentum 
the two different helicity curves coincide, while at $|\vec p|=500\;$MeV there is 
a clear distinction.
} }
\end{figure}

In Fig.~\ref{fig1} we show the spectral functions 
of the helicity-1/2 and helicity-3/2 isobar states, i.e.\ 
$-\text{Im}\,G^{(P)}_{[55]}/\pi$ and $-\text{Im}\,G^{(Q)}_{[11]}/\pi$, in
isospin-symmetric nuclear matter
with Fermi momentum $k_F=270\;$MeV, at momentum 0 and 500 MeV. 
For comparison we also show the results of the non-relativistic calculation of 
Ref.\ \cite{korpa95}, when the isobar propagator has only one component.

The spectral functions of all the other componenets are significantly smaller
than the ones shown in Fig.~\ref{fig1} and for the considered density and for
isobar momentum less than 1 GeV they do not exceed few percent of the dominant spectral
functions, if considering the resonance region. This implies that we can obtain 
practically the same results by 
calculating only two components of the propagator and that these can be obtained
from only two components of the self energy. In vacuum it suffices to calculate
only one component (say $\sigma^{(Q)}_{[11]}$) of the self energy and then the
corresponding propagator to perform a relativistic treatment with  
few-percent accuracy.

We now want to clarify the content of the schemes used in Refs.~\cite{xia94,vandaele02},
where the assumption is made that the in-medium isobar propagator is 
proportional to the spin-3/2 projector. The latter can be written through
the terms of our basis introduced in Eqs.~(\ref{qs}) and (\ref{ps}) as:
\begin{equation}
P_{3/2}^{\mu\nu}=Q^{\mu\nu}_{[11]}+Q^{\mu\nu}_{[22]}+P^{\mu\nu}_{[55]}
+P^{\mu\nu}_{[66]},
\label{spinproj}
\end{equation}
corresponding to positive-energy and negative-energy helicity-3/2 and 
helicity-1/2 terms of the spin-3/2 sector. If one multiplies Eq.~(\ref{spinproj})
with $a\,p\hspace{-1.7mm}/ +b\,\openone$ (on the
left or the right) to get the complete expression for the propagator
(as done in Refs.~\cite{xia94,vandaele02}), only the positive-energy
and negative-energy content will change, while the coefficients
giving the helicity-1/2 and helicity-3/2 components necessarily remain
the same. In this way the splitting between different helicity states
of the spin-3/2 sector is neglected in the approach mentioned, which
is in this respect reminiscent of the non-relativistic treatment
(however, keeping both positive-energy and negative-energy
contributions). We stress that this is not equivalent (in the medium)
to using the general form of the propagator and then applying the
projector (\ref{spinproj}), which would take into account the 
helicity splitting.

\section{Nuclear photoabsorption in the isobar region}
\label{photo}
We now turn to the photon absorption cross section on a
nucleon in the isobar region, i.e.\ for photon energies 
between 0.2 and 0.5 GeV. First we consider the free nucleon case and
subsequently that of a large nucleus, whose nucleons and the isobar are modelled
by their nuclear matter properties. 

We start with 
the absorption of the photon by a free nucleon. As a consequence of the
unitarity of the $S$ matrix the total photon-absorption cross section
on the nucleon
is proportional to the imaginary part of the photon-nucleon forward-scattering
amplitude,
\be
\sigma_T=\frac{1}{2M_N q_0}\,\text{Im}\,A_{\gamma\text{N}},
\ee
where we work in the rest frame of the nucleon and $q_0$ is the photon energy 
in that frame.

For the $\gamma\text{N}\Delta$ vertex we use the dominant magnetic
dipole term \cite{pascalutsa03}:
\be
\mathcal{L}_{\gamma\text{N}\Delta}=\frac{3e}{2M_N(M_N+M_\Delta)}ig_m \bar{N}
T_3^\dagger \tilde{F}^{\mu\nu}\partial_\mu\Delta_\nu+\text{h.c.},\label{gnd}
\ee
where $\tilde F_{\mu\nu}\equiv \varepsilon_{\mu\nu\alpha\beta}\,F^{\alpha\beta}/2$
and $F_{\mu\nu}\equiv \partial_\mu A_\nu-\partial_\nu A_\mu$, with $A_\mu$ being
the electromagnetic field and $g_m$ a dimensionless number giving the strength of
the transition. The photon-nucleon forward-scattering amplitude through 
the isobar intermediate state for arbitrary $q$ photon 4-momentum and $k$ nucleon
4-momentum becomes:
\be
A_{\gamma\text{N}}(q,k)=\frac{2}{3}\;\frac{1}{4}\;\text{Tr}\,
\left[(k\hspace{-1.7mm}/ +M_N)\,\Gamma^\mu_{\;\;\alpha}(k,q)\, G^{\alpha\beta}(q+k)\, 
\Gamma_{\beta\mu}(k,q)\right],
\ee
where the factor 1/4 comes from averaging over the spin of the nucleon and 
photon polarization, and 
\[
\Gamma_{\mu\nu}(k,q)\equiv \frac{3e}{2M_N(M_N+M_\Delta)}\, g_m\, \varepsilon
_{\mu\nu\alpha\beta} \,k^\alpha \,q^\beta.
\]
Inserting the expansion of the isobar propagator (\ref{expb}) and calculating the
trace we arrive at the forward-scattering amplitude in the form
\be
A_{\gamma\text{N}}(q,k)=\frac{1}{3}\,g_m^2\,h_m^2\,\left(
\sum_{i,j=1}^2 a^{(Q)}_{[ij]}(k,q) G^{(Q)}_{[ij]}(q+k) +
\sum_{i,j=1}^6 a^{(P)}_{[ij]}(k,q) G^{(P)}_{[ij]}(q+k) \right),
\label{gnamp}
\ee 
where
\[
h_m\equiv \frac{3e}{2M_N(M_N+M_\Delta)},
\]
and the expressions for $a^{(Q)}_{[ij]}(k,q)$ and $a^{(P)}_{[ij]}(k,q)$ are 
given in the Appendix \ref{apB}, and $G^{(Q)}_{[ij]}(q+k)$ and $G^{(P)}_{[ij]}(q+k)$
are the expansion coefficients of the isobar propagator, as defined by Eq.~(\ref{expb}).

\begin{figure}
\centering{
\includegraphics[width=0.6\textwidth] {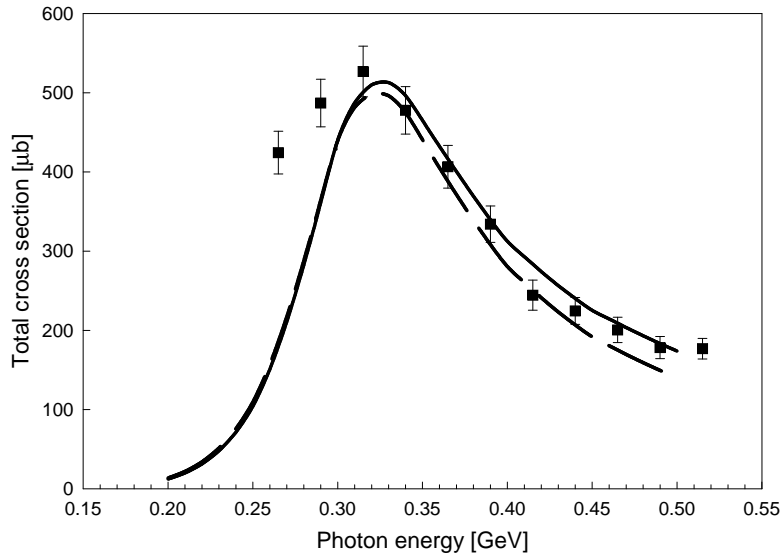}
\caption{\label{fig2}The total photoabsorption cross section on free nucleon, with
the isobar propagator taken from Ref.\ \cite{korpa97}. The calculation leading to
the solid line
is based on magnetic-dipole $\gamma\textrm{N}\Delta$ vertex without form factor,
with coupling $g_m=3$. The dashed line shows the effect of the form factor used
in Ref.~\cite{xia94}.
} }
\end{figure}

The calculated cross section is shown in Fig.~\ref{fig2}, where the dressed isobar 
propagator in vacuum 
is calculated using the free propagators of the nucleon and pion. 
For the  
$\pi\text{N}\Delta$ form factor we used an exponential function of the isobar's
4-momentum $p$:
\begin{equation}
F_{\pi\text{N}\Delta}(p^2)=\exp \left[ -\left(p^2-(m_N+m_\pi)^2\right)/\Lambda^2
\right],
\label{ffdelta}
\end{equation}
which has been fitted to reproduce the pion-nucleon
scattering phase shift in the isobar channel up to pion laboratory momentum of
500 MeV \cite{korpa97}, leading to $\Lambda=0.97\;$GeV and the $\pi\text{N}\Delta$
coupling multiplying Eq.~(\ref{ffdelta}), $g_{\pi\text{N}\Delta}=20.2\;$GeV$^{-1}$. 
The solid line in Fig.~\ref{fig2} is obtained without introducing a form factor
for the $\gamma\text{N}\Delta$ vertex, although the coupling (\ref{gnd}) allows
it without affecting the conservation of the current to which the photon field
couples. The dashed line corresponds to the $\gamma\text{N}\Delta$ form factor
used in Ref.~\cite{xia94}. As one can see in the energy region of interest 
the effect of the form factor is rather small.
The points with error bars show measurement
results from Ref.~\cite{armstrong72}, where we did not attempt to subtract the
background in view of its uncertainty \cite{armstrong72}. 

The only terms giving significant contributions
in the sum of Eq.~(\ref{gnamp}) are those containing $G^{(Q)}_{[11]}(p)$ and 
$G^{(P)}_{[55]}(p)$, where we find numerically that the first one gives three  
times the contribution of
the second. This confirms the interpretation of the $G^{(Q)}_{[11]}(p)$ as corresponding
to helicity 3/2 and $G^{(P)}_{[55]}(p)$ to helicity 1/2, since for a magnetic dipole
transition the $\gamma \text{N}\rightarrow \Delta$ amplitudes in the isobar
helicity basis are related by a factor of $\sqrt{3}$ \cite{jones73}.

We now turn to the nuclear medium. Modification of the following ingredients
of the calculation can be expected: the nucleon spectral function,
the delta spectral function and the $\gamma\text{N}\Delta$ vertex.
For the nucleons we use the Fermi-gas approximation, which means
that we neglect the small broadening of
the nucleon-hole spectral function, but take care of its Fermi motion
which contributes significantly to the broadening of the cross-section energy
distribution.
The small shift of the spectral-function peak due to binding we accomodate
by allowing a mean-field shift of the isobar's mass, since only the difference
of the two values plays a role. The main effect of the medium (apart from the 
nucleon's Fermi motion) comes from
the modified isobar propagator which we compute using an in-medium pion
propagator. To assure a reasonable width of the isobar it is necessary 
to include a rather soft $\pi\text{N}\Delta$ form factor in this calculation,
giving support to previous conclusions about its strong momentum
dependence \cite{koepf96}. For the $\gamma\text{N}\Delta$ vertex we 
assume absence of medium modification.

The pion propagator in the nuclear medium acquires properties distinguishing
it significantly from the vacuum case \cite{oset82,xia94,korpa95,lutz03}. 
The particle-hole excitation
introduces a spectral-function strength at low energy, while the isobar-hole
one gives a finite width to the main peak and gives rise to the isobar-hole
branch at energies above the central maximum (with which it can merge).
The quantitative result for the in-medium propagator depends sensitively
on the pion-nucleon-nucleon and pion-nucleon-isobar form factor used,
as well as on the values of the Migdal's $g'$ parameters ($g'_\text{NN},
g'_{\text{N}\Delta}, g'_{\Delta\Delta}$).  
The low-energy strength in the pion spectral function and the broadening of
the main maximum, usually accompanied with a shift toward smaller energy,
enhance the decay width of the isobar
into a nucleon and pion, while the Pauli blocking decreases the decay probability. 
The last
mechanism is effective only at small isobar momenta. In self-consistent
calculations (or in ones using a dressed pion) this pronounced broadening of the 
isobar can cause problems. In 
Ref.~\cite{xia94}
a large mean-field shift of the isobar energy was used to suppress this
unwelcome result, while in Ref.~\cite{korpa95} the pion-momentum
dependent $\pi\text{N}\Delta$ form factor achieved the same effect.

Our point of view is that it is plausible to use a suitable form factor which suppresses the far
off-shell (with energy around zero and momentum of few hundred MeV)
pion contribution to the isobar self energy and in the following we applied 
this strategy. 
This is corroborated by considerations of deep inelastic scattering on nucleon 
in the pion-cloud model \cite{koepf96}, where the pion can be far off-shell, and
soft $\pi\text{NN}$ and $\pi\text{N}\Delta$ form factors prevent overestimation of antiquark
distributions.
To be consistent with the vacuum calculation of the isobar
self energy, we introduce an (exponential) form factor depending on the pion 4-momentum
squared (which reduces to unity for an on-shell pion):
\be
f_{\pi\text{N}\Delta}(q^2_\pi)=\exp \left[ -(q^2_\pi-m_\pi^2)^2/
\Lambda_{\pi\text{N}\Delta}^4\right].
\label{ffpi}
\ee
Expression (\ref{ffpi}) multiplyes Eq.~(\ref{ffdelta}) to give the full
$\pi\text{N}\Delta$ form factor:
\be
\tilde{F}_{\pi\text{N}\Delta}(p^2,q_\pi^2)=F_{\pi\text{N}\Delta}(p^2)
\cdot f_{\pi\text{N}\Delta}(q^2_\pi).
\ee
We remark that a hard form factor depending on $q^2_\pi$ was introduced also in 
Ref.~\cite{xia94}, but it did not provide a significant suppression of the
off-shell pion contribution. The parameter $\Lambda_{\pi\text{N}\Delta}$ is 
expected to be of the order of a few $m_\pi$.

In order to be able to take into account the Fermi motion of the nucleons we start with
the photon-absorption cross section on a nucleon moving with momentum $\vec k$:
\be
\sigma_T(k,q)=\frac{1}{2 q_0 \left[ E_N(\vec k)-|\vec k| \cos \theta \right]}
\,\text{Im}\,A_{\gamma\text{N}}(q,k),
\label{stgen}
\ee
where $q$ is the photon's 4-momentum and $\theta$ is the angle between the 
photon and nucleon 3-momenta. The expression 
(\ref{gnamp}) for the forward-scattering amplitude and the coefficient functions
$a^{(Q)}_{[ij]}(k,q)$ and $a^{(P)}_{[ij]}(k,q)$ are general and apply also in 
this case. The numerical computation is performed in the rest frame of the medium
and involves averaging over the nucleon's momentum:
\be
\sigma_T(q_0)=\frac{4}{\rho}\,\int_0^{k_F} \frac{k^2 dk}{(2\pi)^2} 
\int_{-1}^1 d\mu\, \sigma_T(q,k),
\ee
with $\rho=2k_F^3/3\pi^2$.

\begin{figure}
\centering{
\includegraphics[width=0.6\textwidth] {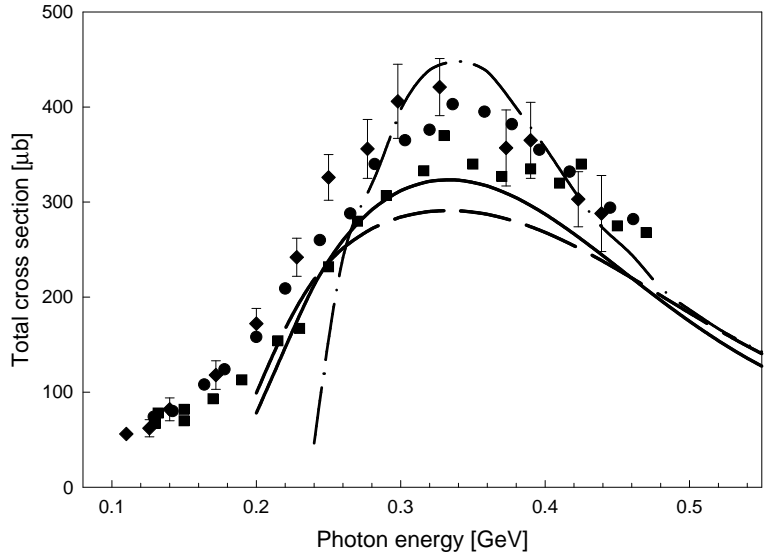}
\caption{\label{fig3}The total photoabsorption cross section on an
in-medium nucleon at saturation density. 
The dash-dot line is based on the free pion propagator. The other two
curves have been
calculated using the pion propagator from Ref.~\cite{korpa03} with
$g'_{\text{NN}}=0.8, g'_{\text{N}\Delta}=g'_{\Delta\Delta}=0.6$. 
For the solid line the form-factor cutoffs are:
$\Lambda_{\pi\text{NN}}=0.5\;$GeV, $\Lambda_{\pi\text{N}\Delta}=0.7\;$GeV, while  
for the dashed line $\Lambda_{\pi\text{NN}}=0.5\;$GeV, 
$\Lambda_{\pi\text{N}\Delta}=0.8\;$GeV. When using the dressed pion 
propagator a mean-field shift of $-30\;$MeV has been introduced for the
isobar mass.
No form factor has been used for the $\gamma\text{N}\Delta$
vertex. The experimental points are from Ref.~\cite{ahrens84} for uranium
and lead.
} }
\end{figure}

In Fig.~\ref{fig3} we show the calculated results, solid, dashed and dot-dash lines,
compared to experimental points from Ref.~\cite{ahrens84} for uranium and lead
nuclei. 
The dash-dot line is obtained using a free pion propagator, in which case
we observe that the isobar becomes too narrow to describe the width of the
measured data, even with included 
smearing due to Fermi motion of the nucleons.

For the two other curves the dressed pion
propagator was taken from the result of a self-consistent calculation based on
the pion-nucleon scattering amplitude, Ref.~\cite{korpa03}. We also
checked that using a pion propagator from some other computations,
i.e.\ from Ref.~\cite{korpa95} and Ref.~\cite{lutz03} does not produce
significantly different 
results (with small adjustments of form-factor cutoffs). 
Actually, using a non self-consistently calculated pion propagator
needs a softer form factor (\ref{ffpi}) because of the more pronounced strength
of the particle-hole branch.
The mentioned latter two curves, solid and dashed line,
show the sensitivity to the $\Lambda_{\pi\text{N}\Delta}$ cut-off in 
the $\pi\text{N}\Delta$
vertex, the solid line corresponding to 
$\Lambda_{\pi\text{N}\Delta}=0.7\;$GeV and the 
dashed line to $\Lambda_{\pi\text{N}\Delta}=0.8\;$GeV. 
In both cases a $\pi\text{NN}$ form factor with $\Lambda_{\pi\text{NN}}=0.5\;$GeV
was used and 
we introduced
a mean-field shift of the isobar mass (relative to that of the nucleon)
of $-30\;$MeV, to better match the position of the experimental peak.
The need for very soft $\pi\text{NN}$ form factor we attribute to the
properties of the calculational scheme of Ref.~\cite{korpa03}, in which
case no form factor was used in the computation of the loop integrals,
determining the self energy of nucleon excitations.
We did not include 
a form factor for the $\gamma\text{N}\Delta$ vertex, which would
affect little the basic shape of the curves, only introducing some suppression
at photon energies larger than 0.3 GeV.
Subtraction of the background of the experimental points was not attempted. 
The relative importance of the background is expected to increase with
leaving the central part of the isobar resonance.

\begin{figure}
\centering{
\includegraphics[width=0.6\textwidth] {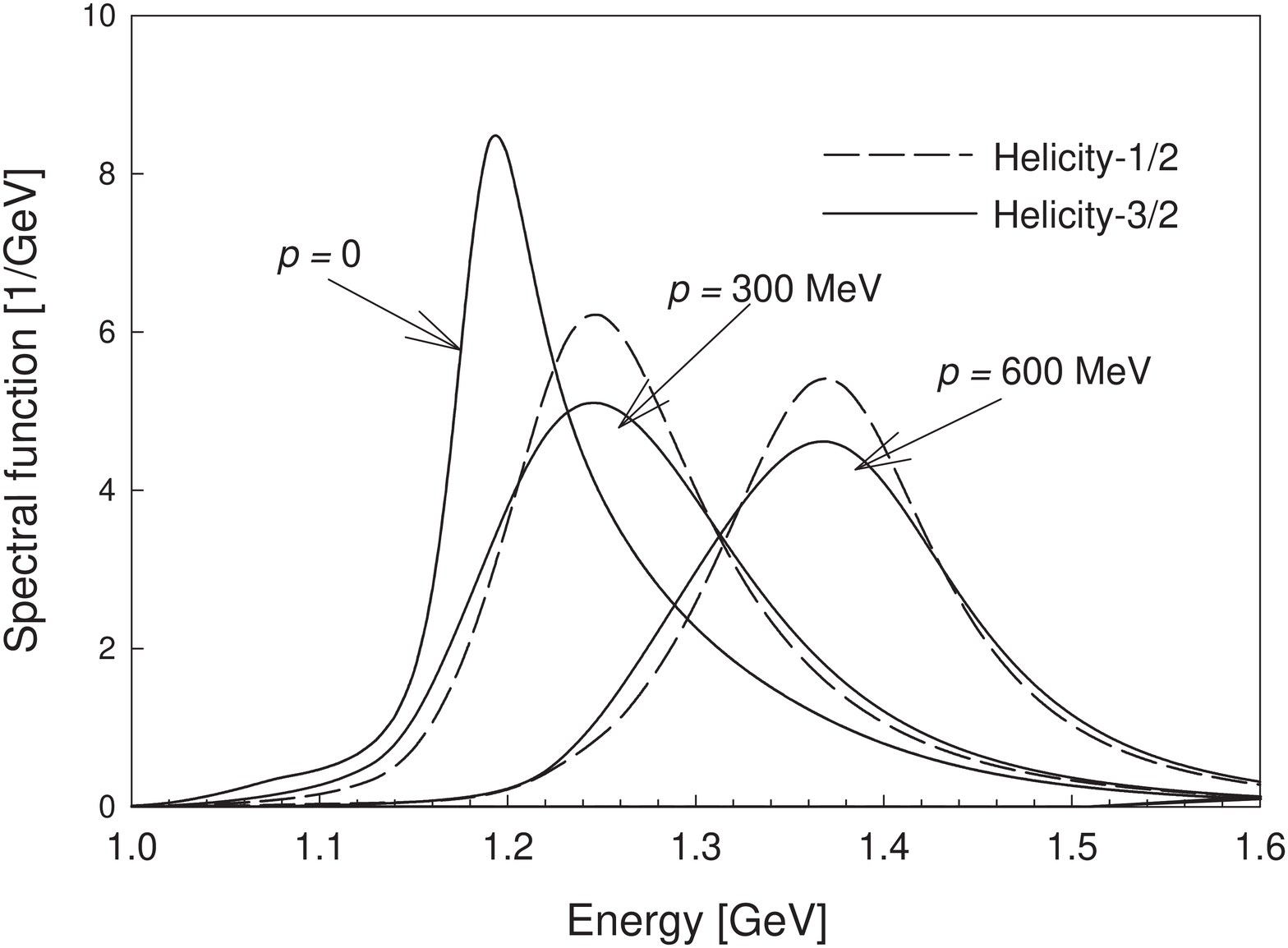}
\caption{\label{fig4}The isobar spectral function in the medium, calculated 
using the pion propagator from Ref.~\cite{korpa03} and with
$\Lambda_\pi=0.6\;$GeV (leading to the solid line in Fig.~\ref{fig3}). 
The dashed line  shows the helicity-1/2 and the
solid line the helicity-3/2 component. The three groups of curves, from
left to right, correspond to momenta $|\vec p|=0$, $|\vec p|=300\;$MeV 
and $|\vec p|=600\;$MeV.
} }
\end{figure}

In Fig.~\ref{fig4} we show the isobar spectral function (helicity-1/2 and
helicity-3/2 components) in the nuclear medium for different momenta, for
parameter values leading to the solid line in Fig.~\ref{fig3}. 
We observe a non-negligible splitting between the two helicity
states and the width is, though increasing with momentum, not very different from
that in vacuum. The positions of the maxima are shifted with respect to the 
vacuum case by slightly less than the introduced $-30\;$MeV mean-field shift.
More detailed examination of the momentum dependence of the two helicities 
shows that their splitting is relatively small below $|\vec p|=200\;$MeV,
reaches maximum around $300\;$MeV and then decreases with increasing
momentum, becoming again small above $900\;$MeV.

\section{Conclusion}
We studied the in-medium behavior of the 
isobar, using a relativistically covariant approach for
describing its self energy and propagator. A convenient basis is
provided by the subset of terms used for the in-medium
generalization of the partial-wave expansion introduced in 
Ref.~\cite{lutz02}. Using a generalized pion-nucleon-isobar vertex that allows
for spin-1/2 components we observed a
small (on the order of a few percent of the spin-3/2 spectral function) 
presence of spin-1/2 components both
in vacuum and in nuclear matter of saturation density. In the nuclear 
medium the two helicity states appear with different spectral functions at 
nonzero isobar momentum (in the rest frame of the medium). 

The dominance of the spin-3/2 states allows a much simplified,
approximate calculation (not used in the present case) with only one 
component in the vacuum (when the
two helicity states are degenerate) and two components in the medium
with high precision, if the density of the medium does not significantly
exceed the saturation density. Actually it suffices to calculate
the relevant one or two components of the self energy to obtain the appropriate
components of the dressed propagator. Physical quantities are expressed 
in terms of these propagator components. These considerations apply to
the energy region where the isobar resonance is prominent, further off shell
the non-spin-3/2 terms in the propagator may become sizable compared to the
spin-3/2 components.

Computation of the total photo-absorption cross section on the free and
in-medium nucleon in the isobar region shows reasonable agreement with the
data, even without introducing a form factor for the photon-nucleon-isobar
vertex. In the medium one has to take into account the dressing of the 
pion propagator, which leads to an additional broadening of the isobar,
mainly due to the particle-hole excitation. This broadening is reduced
by the use of an off-shell pion-nucleon-isobar form factor.
The required
cut-off for the exponential form of (\ref{ffpi}) turns out be 
5$-$6$\; m_\pi$. 

\begin{acknowledgments} 
This work is part of the research program of the ``Stichting voor
Fundamenteel Onderzoek der Materie" (FOM) with financial support
from the ``Nederlandse Organisatie voor Wetenschappelijk
Onderzoek" (NWO). 
C.L.K would like to thank the NWO for providing a 
visitors stipend and the K.V.I. (Groningen) for the kind hospitality.
The authors would like to thank 
Olaf Scholten for useful discussions.
\end{acknowledgments}

\appendix
\section{}
\label{apA}
The nonzero expansion coefficients of the inverse of the free isobar propagator,
Eq.\ (\ref{freeprop}), are given for $p^2>0$ by:
\begin{eqnarray}
g^{(Q)-1}_{[11]}=&\sqrt{p^2}-M, \qquad &g^{(Q)-1}_{[22]}=-(\sqrt{p^2}+M),\nonumber\\
g^{(P)-1}_{[11]}=&2(\sqrt{p^2}+M), &g^{(P)-1}_{[13]}=\sqrt{3}M\nonumber\\
g^{(P)-1}_{[22]}=&-2(\sqrt{p^2}-M), &g^{(P)-1}_{[24]}=-\sqrt{3}M\nonumber\\
g^{(P)-1}_{[31]}=&\sqrt{3}M, &g^{(P)-1}_{[42]}=-\sqrt{3}M\nonumber\\
g^{(P)-1}_{[55]}=&\sqrt{p^2}-M, &g^{(P)-1}_{[66]}=-(\sqrt{p^2}+M),
\end{eqnarray}
with $M$ being the bare mass of the isobar.

The expansion coefficients, $\sigma^{(Q)}_{[ij]}$ and $\sigma^{(P)}_{[ij]}$,
of the isobar self energy 
form a symmetric 2$\times$2 and a symmetric 6$\times$6 matrix, with
entries given for $i\le j$ by:
\begin{eqnarray}
&&\sigma^{(Q)}_{[11]}=M_N L_3+L_7,\nonumber\\
&&\sigma^{(Q)}_{[12]}=-i L_8,\nonumber\\
&&\sigma^{(Q)}_{[22]}=M_N L_3-L_7,\nonumber\\
&&\sigma^{(P)}_{[11]}=\frac{1}{3}\left[ M_N(2L_3-L_5)-2L_7+L_{12}\right],\nonumber\\
&&\sigma^{(P)}_{[12]}=\frac{i}{3}\left( -2L_8+L_{11}\right),\nonumber\\
&&\sigma^{(P)}_{[13]}=\frac{1}{\sqrt{3}}\left[-\sqrt{p^2}(2L_3-L_5)+2L_7-L_{12}\right],
\nonumber\\
&&\sigma^{(P)}_{[14]}=\frac{i}{\sqrt{3}}\left[M_N(\sqrt{p^2}L_2-L_6)-\sqrt{p^2}L_6+%
L_{10}\right],
\nonumber\\
&&\sigma^{(P)}_{[15]}=\frac{i\sqrt{2}}{3}\left(2L_8-L_{11}\right),\nonumber\\
&&\sigma^{(P)}_{[16]}=\frac{\sqrt{2}}{3}\left[M_N(L_3+L_5)-L_7-L_{12}\right],\nonumber\\
&&\sigma^{(P)}_{[22]}=\frac{1}{3}\left[M_N(2L_3-L_5)+2L_7-L_{12}\right],\nonumber\\
&&\sigma^{(P)}_{[23]}=\frac{i}{\sqrt{3}}\left[M_N(\sqrt{p^2}L_2-L_6)-\sqrt{p^2}L_6-%
L_{10}\right],\nonumber\\
&&\sigma^{(P)}_{[24]}=\frac{1}{\sqrt{3}}\left[-\sqrt{p^2}(2L_3-L_5)+2L_7-L_{12}\right],
\nonumber\\
&&\sigma^{(P)}_{[25]}=\frac{\sqrt{2}}{3}\left[M_N(L_3+L_5)+L_7+L_{12}\right],
\nonumber\\
&&\sigma^{(P)}_{[26]}=\frac{i\sqrt{2}}{3}\left(2L_8-L_{11}\right),
\nonumber
\end{eqnarray}
\begin{eqnarray}
&&\sigma^{(P)}_{[33]}=M_N p^2 L_0+(p^2-2M_N\sqrt{p^2})L_1+(M_N-2\sqrt{p^2})L_4
+L_9,\nonumber\\
&&\sigma^{(P)}_{[34]}=i(-p^2 L_2+2\sqrt{p^2}L_6-L_{10}),\nonumber\\
&&\sigma^{(P)}_{[35]}=i\sqrt{\frac{2}{3}}\left[ -M_N\sqrt{p^2} L_2+
(M_N-\sqrt{p^2})L_6+L_{10}\right],\nonumber\\
&&\sigma^{(P)}_{[36]}=\sqrt{\frac{2}{3}}\left[ -\sqrt{p^2} (L_3+L_5)
+L_7+L_{12}\right],\nonumber\\
&&\sigma^{(P)}_{[44]}=M_N p^2 L_0-(p^2+2M_N \sqrt{p^2})L_1+(M_N+2\sqrt{p^2})L_4
-L_9,\nonumber\\
&&\sigma^{(P)}_{[45]}=\sqrt{\frac{2}{3}} \left[ -\sqrt{p^2}(L_3+L_5)+L_7+L_{12}\right],
\nonumber\\
&&\sigma^{(P)}_{[46]}=i\sqrt{\frac{2}{3}} \left[-M_N \sqrt{p^2} L_2+(M_N+\sqrt{p^2})
L_6-L_{10}\right],\nonumber\\
&&\sigma^{(P)}_{[55]}=\frac{1}{3}\left[M_N(L_3-2L_5)+L_7-2L_{12}\right],\nonumber\\
&&\sigma^{(P)}_{[56]}=\frac{i}{3}\left( 5L_8+2L_{11}\right),\nonumber\\
&&\sigma^{(P)}_{[66]}=\frac{1}{3}\left[M_N(L_3-2L_5)-L_7+2L_{12}\right].
\label{sigmaexp}
\end{eqnarray}

The pion-nucleon loop integrals $L_i$ ($i=0,\cdots,12$) are regularized by the
form factor present in the pion-nucleon-isobar vertex.
The imaginary part of the loop integrals is first computed
from the imaginary parts of the nucleon and pion propagator and with the help
of the form factor made to approach zero at large energy. The real part is then 
computed from a convergent dispersion integral. For the
nucleons we use a Fermi-gas description, while the pion propagator can be any
result, obtained in an independent calculation. 

The imaginary part of the loop integrals is then given by the following expression:
\be
\textrm{Im}\, L_i(p,u)=\frac{g^2_{\pi\textrm{N}\Delta}}{8\pi^2}
\int_{k_F}^\infty \frac{k^2 dk}{E_k}\int_{-1}^1 d\mu \,
F(p,k)^2\; \textrm{Im}\left[ D_\pi(p_0-E_k,|\vec p - \vec k|)\right]\, K_i(p,k),
\ee
where $F(p,k)$ is the $\pi\textrm{N}\Delta$ form factor, 
$\mu\equiv \cos\theta(\vec p,\vec k)$, $D_\pi$ the pion propagator 
and the functions $K_i(p,k)$ are:
\begin{eqnarray} 
&&\begin{array}{ccc}
K_0=1,\;K_1=k\cdot \hat p,& K_2=-X\cdot k,&
K_3=\frac{1}{2}\left[M_N^2-(k\cdot \hat p) ^2+(X\cdot k)^2\right],\\
K_4=(k\cdot \hat p)^2,& K_5=(X\cdot k)^2 &
K_6=-(X\cdot k) (k\cdot \hat p),\\ 
K_7=(k\cdot \hat p) K_3, &
K_8=-(X\cdot k) K_3, & 
K_9=(k\cdot \hat p)^3, \\ 
K_{10}=-(X\cdot k) (k\cdot \hat p)^2, &
K_{11}=-(X\cdot k)^3, & K_{12}=(k\cdot \hat p) (X\cdot k)^2,
\end{array}
\end{eqnarray}
with $k\cdot\hat p\equiv k\cdot p/\sqrt{p^2}$ and $X\equiv X(p,u)$ defined
in Eq.~(\ref{def-basic}).

\section{}
\label{apB}
The terms $a^{(Q)}_{[ij]}(k,q)$ and $a^{(P)}_{[ij]}(k,q)$ in
Eq.~(\ref{gnamp}) are symmetric under the exchange of $i$ and $j$, thus we give
them only for $i\le j$. 
Defining $M_\pm\equiv M_N \pm k\cdot p/\sqrt{p^2}$ and writing simply $X_\mu$
for $X_\mu(p,u)$, the nonzero values are as follows:
\begin{eqnarray}
a^{(Q)}_{11} &=& \frac{p^2}{2}M_+\left[ -M_+ M_- +\left( X\cdot k\right)^2\right],
\nonumber\\
a^{(Q)}_{12} &=& \frac{-i p^2}{2}\left( X\cdot k\right)
\left[ M_+ M_- -\left( X\cdot k\right)^2\right],  \nonumber\\
a^{(Q)}_{22} &=& \frac{p^2}{2}M_-\left[ -M_+ M_- +\left( X\cdot k\right)^2\right],
\nonumber\\
a^{(P)}_{11} &=& \frac{-2p^2}{3}M_+M_-^2,
\nonumber
\end{eqnarray}
\begin{eqnarray}
a^{(P)}_{12} &=& \frac{2ip^2}{3}M_+M_- \left( X\cdot k\right),\nonumber\\
a^{(P)}_{15} &=& \frac{i\sqrt{2}p^2}{3}M_+M_- \left( X\cdot k\right),\nonumber\\
a^{(P)}_{16} &=& \frac{\sqrt{2}p^2}{2}M_- \left[ \frac{1}{3}M_+ M_-+\left( X\cdot k\right)^2
\right],\nonumber\\
a^{(P)}_{22} &=& \frac{-2p^2}{3}M_+^2M_-,
\nonumber\\
a^{(P)}_{25} &=& \frac{\sqrt{2}p^2}{2}M_+ \left[ \frac{1}{3}M_+ M_-+\left( X\cdot k\right)^2
\right],\nonumber\\
a^{(P)}_{26} &=& \frac{i\sqrt{2}p^2}{3}M_+M_- \left( X\cdot k\right),\nonumber\\
a^{(P)}_{55} &=& \frac{-p^2}{2}M_+ \left[ \frac{5}{3}M_+ M_-+\left( X\cdot k\right)^2
\right],\nonumber\\
a^{(P)}_{56} &=& -i\frac{(X\cdot k)}{2}p^2 \left[ \frac{7}{3}M_+ M_-+3\left( X\cdot k\right)^2
\right],\nonumber\\
a^{(P)}_{66} &=& \frac{-p^2}{2}M_- \left[ \frac{5}{3}M_+ M_-+\left( X\cdot k\right)^2
\right].
\label{as}
\end{eqnarray}
We note that all terms containing index 3 or 4 are identically zero.


\begin{thebibliography}{20}
\expandafter\ifx\csname natexlab\endcsname\relax\def\natexlab#1{#1}\fi
\expandafter\ifx\csname bibnamefont\endcsname\relax
  \def\bibnamefont#1{#1}\fi
\expandafter\ifx\csname bibfnamefont\endcsname\relax
  \def\bibfnamefont#1{#1}\fi
\expandafter\ifx\csname citenamefont\endcsname\relax
  \def\citenamefont#1{#1}\fi
\expandafter\ifx\csname url\endcsname\relax
  \def\url#1{\texttt{#1}}\fi
\expandafter\ifx\csname urlprefix\endcsname\relax\def\urlprefix{URL }\fi
\providecommand{\bibinfo}[2]{#2}
\providecommand{\eprint}[2][]{\url{#2}}

\bibitem[{\citenamefont{Rarita and Schwinger}(1941)}]{rarita41}
\bibinfo{author}{\bibfnamefont{W.}~\bibnamefont{Rarita}} \bibnamefont{and}
  \bibinfo{author}{\bibfnamefont{J.}~\bibnamefont{Schwinger}},
  \bibinfo{journal}{Phys.\ Rev.} \textbf{\bibinfo{volume}{60}},
  \bibinfo{pages}{61} (\bibinfo{year}{1941}).

\bibitem[{\citenamefont{Davidson et~al.}(1991)\citenamefont{Davidson,
  Mukhopadhyay, and Wittmann}}]{davidson91}
\bibinfo{author}{\bibfnamefont{R.~M.} \bibnamefont{Davidson}},
  \bibinfo{author}{\bibfnamefont{N.~C.} \bibnamefont{Mukhopadhyay}},
  \bibnamefont{and} \bibinfo{author}{\bibfnamefont{R.~S.}
  \bibnamefont{Wittmann}}, \bibinfo{journal}{Phys.\ Rev.\ D}
  \textbf{\bibinfo{volume}{43}}, \bibinfo{pages}{71} (\bibinfo{year}{1991}).

\bibitem[{\citenamefont{Pascalutsa}(1998)}]{pascalutsa98}
\bibinfo{author}{\bibfnamefont{V.}~\bibnamefont{Pascalutsa}},
  \bibinfo{journal}{Phys.\ Rev.\ D} \textbf{\bibinfo{volume}{58}},
  \bibinfo{pages}{096002} (\bibinfo{year}{1998}).

\bibitem[{\citenamefont{Pascalutsa and Timmermans}(1999)}]{pascalutsa99}
\bibinfo{author}{\bibfnamefont{V.}~\bibnamefont{Pascalutsa}} \bibnamefont{and}
  \bibinfo{author}{\bibfnamefont{R.}~\bibnamefont{Timmermans}},
  \bibinfo{journal}{Phys.\ Rev.\ C} \textbf{\bibinfo{volume}{60}},
  \bibinfo{pages}{042201} (\bibinfo{year}{1999}).

\bibitem[{\citenamefont{Pascalutsa}(2001)}]{pascalutsa01}
\bibinfo{author}{\bibfnamefont{V.}~\bibnamefont{Pascalutsa}},
  \bibinfo{journal}{Phys.\ Lett.\ B} \textbf{\bibinfo{volume}{503}},
  \bibinfo{pages}{85} (\bibinfo{year}{2001}).

\bibitem[{\citenamefont{Korpa and Malfliet}(1995)}]{korpa95}
\bibinfo{author}{\bibfnamefont{C.~L.} \bibnamefont{Korpa}} \bibnamefont{and}
  \bibinfo{author}{\bibfnamefont{R.}~\bibnamefont{Malfliet}},
  \bibinfo{journal}{Phys.\ Rev.\ C} \textbf{\bibinfo{volume}{52}},
  \bibinfo{pages}{2756} (\bibinfo{year}{1995}).

\bibitem[{\citenamefont{de~Jong and Malfliet}(1992)}]{dejong92}
\bibinfo{author}{\bibfnamefont{F.}~\bibnamefont{de~Jong}} \bibnamefont{and}
  \bibinfo{author}{\bibfnamefont{R.}~\bibnamefont{Malfliet}},
  \bibinfo{journal}{Phys.\ Rev.\ C} \textbf{\bibinfo{volume}{46}},
  \bibinfo{pages}{2567} (\bibinfo{year}{1992}).

\bibitem[{\citenamefont{Xia et~al.}(1994)\citenamefont{Xia, Siemens, and
  Soyeur}}]{xia94}
\bibinfo{author}{\bibfnamefont{L.}~\bibnamefont{Xia}},
  \bibinfo{author}{\bibfnamefont{P.~J.} \bibnamefont{Siemens}},
  \bibnamefont{and} \bibinfo{author}{\bibfnamefont{M.}~\bibnamefont{Soyeur}},
  \bibinfo{journal}{Nucl.\ Phys.\ A} \textbf{\bibinfo{volume}{578}},
  \bibinfo{pages}{493} (\bibinfo{year}{1994}).

\bibitem[{\citenamefont{Daele et~al.}(2002)\citenamefont{Daele, Korchin, Neck,
  Scholten, and Waroquier}}]{vandaele02}
\bibinfo{author}{\bibfnamefont{L.~V.} \bibnamefont{Daele}},
  \bibinfo{author}{\bibfnamefont{A.~Y.} \bibnamefont{Korchin}},
  \bibinfo{author}{\bibfnamefont{D.~V.} \bibnamefont{Neck}},
  \bibinfo{author}{\bibfnamefont{O.}~\bibnamefont{Scholten}}, \bibnamefont{and}
  \bibinfo{author}{\bibfnamefont{M.}~\bibnamefont{Waroquier}},
  \bibinfo{journal}{Phys.\ Rev.\ C} \textbf{\bibinfo{volume}{65}},
  \bibinfo{pages}{014613} (\bibinfo{year}{2002}).

\bibitem[{\citenamefont{Lutz and Korpa}(2002)}]{lutz02}
\bibinfo{author}{\bibfnamefont{M.~F.~M.} \bibnamefont{Lutz}} \bibnamefont{and}
  \bibinfo{author}{\bibfnamefont{C.~L.} \bibnamefont{Korpa}},
  \bibinfo{journal}{Nucl.\ Phys.\ A} \textbf{\bibinfo{volume}{700}},
  \bibinfo{pages}{309} (\bibinfo{year}{2002}).

\bibitem[{\citenamefont{Korpa}(1997)}]{korpa97}
\bibinfo{author}{\bibfnamefont{C.~L.} \bibnamefont{Korpa}},
  \bibinfo{journal}{Heavy Ion Phys.} \textbf{\bibinfo{volume}{5}},
  \bibinfo{pages}{77} (\bibinfo{year}{1997}).

\bibitem[{\citenamefont{Kaloshin and Lomov}(2004)}]{kaloshin04}
\bibinfo{author}{\bibfnamefont{A.~E.} \bibnamefont{Kaloshin}} \bibnamefont{and}
  \bibinfo{author}{\bibfnamefont{V.~P.} \bibnamefont{Lomov}},
  \bibinfo{journal}{Mod.\ Phys.\ Lett.\ A} \textbf{\bibinfo{volume}{19}},
  \bibinfo{pages}{135} (\bibinfo{year}{2004}).

\bibitem[{\citenamefont{Pascalutsa and Phillips}(2003)}]{pascalutsa03}
\bibinfo{author}{\bibfnamefont{V.}~\bibnamefont{Pascalutsa}} \bibnamefont{and}
  \bibinfo{author}{\bibfnamefont{D.~R.} \bibnamefont{Phillips}},
  \bibinfo{journal}{Phys.\ Rev.\ C} \textbf{\bibinfo{volume}{67}},
  \bibinfo{pages}{055202} (\bibinfo{year}{2003}).

\bibitem[{\citenamefont{Armstrong et~al.}(1972)\citenamefont{Armstrong, Hogg,
  Lewis, Robertson, Brookes, Clough, Freeland, Galbraith, King, Rawlinson
  et~al.}}]{armstrong72}
\bibinfo{author}{\bibfnamefont{T.~A.} \bibnamefont{Armstrong}},
  \bibinfo{author}{\bibfnamefont{W.~R.} \bibnamefont{Hogg}},
  \bibinfo{author}{\bibfnamefont{G.~M.} \bibnamefont{Lewis}},
  \bibinfo{author}{\bibfnamefont{A.~W.} \bibnamefont{Robertson}},
  \bibinfo{author}{\bibfnamefont{G.~R.} \bibnamefont{Brookes}},
  \bibinfo{author}{\bibfnamefont{A.~S.} \bibnamefont{Clough}},
  \bibinfo{author}{\bibfnamefont{J.~H.} \bibnamefont{Freeland}},
  \bibinfo{author}{\bibfnamefont{W.}~\bibnamefont{Galbraith}},
  \bibinfo{author}{\bibfnamefont{A.~F.} \bibnamefont{King}},
  \bibinfo{author}{\bibfnamefont{W.~R.} \bibnamefont{Rawlinson}},
  \bibnamefont{et~al.}, \bibinfo{journal}{Phys.\ Rev.\ D}
  \textbf{\bibinfo{volume}{5}}, \bibinfo{pages}{1640} (\bibinfo{year}{1972}).

\bibitem[{\citenamefont{Jones and Scadron}(1973)}]{jones73}
\bibinfo{author}{\bibfnamefont{H.~F.} \bibnamefont{Jones}} \bibnamefont{and}
  \bibinfo{author}{\bibfnamefont{M.~D.} \bibnamefont{Scadron}},
  \bibinfo{journal}{Ann.\ Phys.} \textbf{\bibinfo{volume}{81}},
  \bibinfo{pages}{1} (\bibinfo{year}{1973}).

\bibitem[{\citenamefont{Koepf et~al.}(1996)\citenamefont{Koepf, Frankfurt, and
  Strikman}}]{koepf96}
\bibinfo{author}{\bibfnamefont{W.}~\bibnamefont{Koepf}},
  \bibinfo{author}{\bibfnamefont{L.~L.} \bibnamefont{Frankfurt}},
  \bibnamefont{and} \bibinfo{author}{\bibfnamefont{M.}~\bibnamefont{Strikman}},
  \bibinfo{journal}{Phys.\ Rev.\ D} \textbf{\bibinfo{volume}{53}},
  \bibinfo{pages}{2586} (\bibinfo{year}{1996}).

\bibitem[{\citenamefont{Oset et~al.}(1982)\citenamefont{Oset, Toki, and
  Weise}}]{oset82}
\bibinfo{author}{\bibfnamefont{E.}~\bibnamefont{Oset}},
  \bibinfo{author}{\bibfnamefont{H.}~\bibnamefont{Toki}}, \bibnamefont{and}
  \bibinfo{author}{\bibfnamefont{W.}~\bibnamefont{Weise}},
  \bibinfo{journal}{Phys.\ Rep.} \textbf{\bibinfo{volume}{83}},
  \bibinfo{pages}{281} (\bibinfo{year}{1982}).

\bibitem[{\citenamefont{Lutz}(2003)}]{lutz03}
\bibinfo{author}{\bibfnamefont{M.~F.~M.} \bibnamefont{Lutz}},
  \bibinfo{journal}{Phys.\ Lett.\ B} \textbf{\bibinfo{volume}{552}},
  \bibinfo{pages}{159} (\bibinfo{year}{2003}).

\bibitem[{\citenamefont{Ahrens et~al.}(1984)\citenamefont{Ahrens, Arends,
  Bourgeois, Carlos, Fallou, Floss, Garganne, Huthmacher, Kneissl, Mank
  et~al.}}]{ahrens84}
\bibinfo{author}{\bibfnamefont{J.}~\bibnamefont{Ahrens}},
  \bibinfo{author}{\bibfnamefont{J.}~\bibnamefont{Arends}},
  \bibinfo{author}{\bibfnamefont{P.}~\bibnamefont{Bourgeois}},
  \bibinfo{author}{\bibfnamefont{P.}~\bibnamefont{Carlos}},
  \bibinfo{author}{\bibfnamefont{J.~L.} \bibnamefont{Fallou}},
  \bibinfo{author}{\bibfnamefont{N.}~\bibnamefont{Floss}},
  \bibinfo{author}{\bibfnamefont{P.}~\bibnamefont{Garganne}},
  \bibinfo{author}{\bibfnamefont{S.}~\bibnamefont{Huthmacher}},
  \bibinfo{author}{\bibfnamefont{U.}~\bibnamefont{Kneissl}},
  \bibinfo{author}{\bibfnamefont{G.}~\bibnamefont{Mank}}, \bibnamefont{et~al.},
  \bibinfo{journal}{Phys.\ Lett.\ B} \textbf{\bibinfo{volume}{146}},
  \bibinfo{pages}{303} (\bibinfo{year}{1984}).

\bibitem[{\citenamefont{Korpa and Lutz}()}]{korpa03}
\bibinfo{author}{\bibfnamefont{C.~L.} \bibnamefont{Korpa}} \bibnamefont{and}
  \bibinfo{author}{\bibfnamefont{M.~F.~M.} \bibnamefont{Lutz}},
  \eprint{nucl-th/0306063}.

\end{thebibliography}
\end{document}